\documentclass[aps,prd,12pt,showpacs, citeautoscript,superscriptaddress]{revtex4-1}

\usepackage[T1]{fontenc}	
\usepackage[english]{babel}	
\usepackage[latin1]{inputenc}	

\usepackage[colorlinks=true, pdfstartview=FitV,linkcolor=blue, citecolor=blue, urlcolor=blue]{hyperref}

\usepackage{amsmath}		
\usepackage{amssymb}		
\usepackage{bbm}			

\usepackage{graphicx}		
\usepackage{graphics}		
\DeclareGraphicsExtensions{.png,.pdf}
\usepackage{color}		

\newcommand{\bs}{\boldsymbol}
\newcommand{\bea}{\begin{eqnarray}}
\newcommand{\eea}{\end{eqnarray}}
\newcommand{\be}{\begin{equation}}
\newcommand{\ee}{\end{equation}}

\newcommand{\nuu}{\nu}
\newcommand{\pii}{p}

\begin{document}
\title{Stability of Minkowski space in ghost-free massive gravity theory}

\author{Mikhail~S.~Volkov}
\affiliation{
Laboratoire de Math\'{e}matiques et Physique Th\'{e}orique CNRS-UMR 7350, \\ 
Universit\'{e} de Tours, Parc de Grandmont, 37200 Tours, France
}
\affiliation{
Department of General Relativity and Gravitation, Institute of Physics,\\
Kazan Federal University, Kremlevskaya str. 18, 420008 Kazan, Russia
}

\begin{abstract}
The energy in the 
ghost-free massive gravity theory  
is calculated via explicitly resolving the 
initial value constraints for spherically symmetric 
deformations of flat space. 
It turns out that  the energy is positive in some cases,
but in other cases it can be negative and even unbounded from below.
This could suggest that ghost 
instability is present. However, 
it seems that the negative energy states cannot
communicate with the positive energy sector since the 
corresponding solutions of the constraint
equations are either not globally defined,  
not asymptotically flat, or singular. 
As a result, they cannot describe  initial data for the decay of 
flat space. At the same time, for globally regular and 
asymptotically flat solutions of the constraints the energy  
is always found to be positive. 
All of this suggests that there is a physical sector of the theory where the
energy is positive and the ghost is suppressed, so that the theory is stable. 
The negative energies show up only in disjoint sectors and thus should be harmless.

\end{abstract}


\maketitle

\section{Introduction}
The idea that gravitons could have a tiny mass, 
which would explain the current cosmic acceleration 
\cite{0004-637X-517-2-565,*1538-3881-116-3-1009}, 
has attracted a lot of interest  after the 
discovery of the special massive gravity theory 
by de Rham, Gabadadze, and Tolley (dRGT) 
\cite{deRham:2010kj} 
(see \cite{Hinterbichler:2011tt,*deRham:2014zqa}
for a review).
 This theory contains two Hamiltonian constraints 
 which eliminate 
one propagating degree of freedom (DOF), usually
associated with the ghost \cite{Hassan:2011ea,*Kluson:2012wf,*Comelli:2012vz} 
-- an unphysical mode with 
a negative kinetic energy rendering the theory unstable 
\cite{Boulware:1973my}. The remaining five DOFs behave well 
in special limits; hence they are 
associated with the five polarizations of the massive graviton,
 and the whole theory is referred to as 
ghost free. 

At the same time, nothing guarantees that removing one DOF kills the 
ghost completely. It may be that the 
remaining five DOFs are still 
contaminated with its remnant, 
suppressed in some cases but present otherwise.  
Indeed, such a concern is supported by the observations of certain    
ghost-type features in the theory    
\cite{DeFelice:2012mx,*Fasiello:2013woa,*Chamseddine:2013lid}. 

A good way to see whether the theory is indeed 
ghost free is to compute the energy since, if the energy is positive, 
the ghost is absent. 
The energy can be defined in the standard way within the canonical 
approach \cite{ADM}, but to evaluate it requires resolving  
the constraints, which are   
known, in general, only implicitly. 
Therefore, the aim of this work 
is to evaluate the energy in the spherically symmetric
sector (the $s$ sector), where the constraints can be obtained explicitly 
and, in some cases, resolved.  
The corresponding solutions can be viewed as initial data 
for the Cauchy problem.

It turns out that the energy is positive in some cases,
but in other cases it is negative and  
unbounded from below. This could suggest that the ghost is still  
present in the theory. However, a closer inspection reveals 
that the negative energy states form disjoint branches that cannot
communicate with the positive energy sector. 
Specifically, the corresponding solutions of the constraint
equations are either not globally defined, 
not asymptotically flat, or singular. 
As a result, they cannot describe  initial data for the decay of the 
flat space. At the same time, for globally regular and 
asymptotically flat solutions of the constraints, the energy  
is always found to be positive.

All of this suggests that there is a physical sector of the theory where the
energy is positive and the ghost is suppressed, so that the theory is stable. 
The negative energies show up only in disjoint sectors, so they are harmless.  
This also suggests that the other seemingly unphysical properties of the theory,
like superluminality  
\cite{Deser:2012qx,*Deser:2013rxa,*Deser:2013eua,*Deser:2013qza,*Deser:2013gpa}, 
may perhaps show up only in disjoint sectors, 
in which case they would be harmless as well.

\section{Massive gravity}	 
The theory is defined 
by the action 
\bea                                      \label{1}
&&\frac{1}{M^2_{\rm Pl}} S
=\int \, \sqrt{-{ g}}\left(\frac{1}{2}\,R-{m^2}\, {\cal U}\right)\,d^4x
\equiv \int {\cal L}\, d^4 x\,.
\,
\eea
For the generic massive gravity that 
reduces to the Fierz-Pauli (FP) theory \cite{Fierz:1939ix} 
in the 
weak field limit, the potential ${\cal U}$ is 
\be                            \label{PF}
{\cal U}= \frac18\,
(H^{\mu}_{~\nu}H^{\nu}_{~\mu}-(H^\mu_{~\mu})^2)
+\ldots 
\ee 
Here $H^\mu_{~\nu}=\delta^\mu_\nu
-g^{\mu\alpha}f_{\alpha\nu}$, where $g^{\mu\nu}$ is the inverse of the spacetime
metric, $f_{\mu\nu}$ is the flat metric, and the dots denote terms which are
higher order in $H^\mu_{~\nu}$ and which can be arbitrary. 
A particular choice of these terms determines the dRGT 
theory \cite{deRham:2010kj}, in which case ${\cal U}$ can be expressed as
\bea                          \label{2}
{\cal U}&=&b_0+
b_1\sum_{a}\lambda_a
+b_2\sum_{a<b}\lambda_a\lambda_b \nonumber \\
&+&b_3\sum_{a<b<c}\lambda_a\lambda_b\lambda_c
+b_4\,\lambda_0\lambda_1\lambda_2\lambda_3\,.
\eea
Here $\lambda_a$ are eigenvalues of 
$
\bs{\gamma}^\mu_{~\nu}=
\sqrt{{{g}}^{\mu\alpha}{{f}}_{\alpha\nu}}
$, 
with
 the 
square root
understood in the sense that  
$
\bs{\gamma}^\mu_{~\alpha}\bs{\gamma}^\alpha_{~\nu}
={{g}}^{\mu\alpha}{{f}}_{\alpha\nu}.
$ 
If the bare cosmological term
is absent, the flat space is a solution of the theory,
and $m$ in \eqref{1} is the FP mass of the graviton 
in the weak field limit, 
then the parameters $b_k$ are expressed 
in terms of two arbitrary $c_3,c_4$ 
as    
$b_0=4c_3+c_4-6$,
$b_1=3-3c_3-c_4$,
$b_2=2c_3+c_4-1$,
$b_3=-c_3-c_4$, and 
$b_4=c_4$.

\section{Hamiltonian in the $S$ sector}
Assuming the spherical coordinates 
$x^\mu=(t,r,\vartheta,\varphi)$, 
the two metrics can be parametrized as 
\bea                              \label{sphere-flat}
ds_g^2&=&-N^2dt^2+\frac{1}{\Delta^2}(dr+\beta\, dt)^2
+R^2\, (d\vartheta^2+\sin^2\vartheta d\varphi^2)  ,
\label{sphere}\nonumber  \\
ds_f^2&=&-dt^2+dr^2+r^2\, (d\vartheta^2+\sin^2\vartheta d\varphi^2),  
\eea 
where $N,\beta,\Delta$, and $R$ depend on $t$ and $r$.
Defining the canonical momenta 
$\pii_\Delta={\partial \cal L}/{\partial \dot{\Delta}}$ and 
$\pii_R={\partial \cal L}/{\partial \dot{R}}$ and  introducing 
${\cal V}=\sqrt{-g}\,{\cal U}$,
the Hamiltonian density is 
\bea              \label{Hs}
{\cal H}&=&\dot{\Delta}\pi_\Delta+\dot{R}\pi_R-{\cal L} 
=N{\cal H}_0+\beta{\cal H}_r+{m^2}{\cal V},~~~
\eea
where the total derivative has been omitted, and  
\bea                  \label{H0r}
{\cal H}_0&=&\frac{\Delta^3}{4R^2}\,\pii_\Delta^2
+\frac{\Delta^2}{2R}\,\pii_\Delta\pii_R 
+\Delta R^{\prime 2}+2R(\Delta R^\prime)^\prime
-\frac{1}{\Delta}\,,~~~~~\nonumber \\
{\cal H}_r&=&\Delta\pii_\Delta^\prime+2\Delta^\prime\pii_\Delta
+R^\prime \pii_R\,.
\eea
These expressions were studied previously in 
\cite{Unruh:1976db,*Kuchar:1994zk}. 
The phase space is spanned by four variables  
$(q^i,p_k)\equiv\{\Delta,R,\pii_\Delta,\pii_R\}$, 
while $N$ and $\beta$ are nondynamical 
since their momenta vanish.

General Relativity is recovered for $m=0$,
in which case varying ${\cal H}$ with respect to 
$N$ and $\beta$ gives 2 constraints: 
${\cal H}_0=0$ and  ${\cal H}_r=0$. 
They are first class and 
generate diffeomorphisms,
which can be used to impose two gauge conditions. 
As a result, there remain 
$4-2-2=0$ independent phase space variables; 
therefore  there is no dynamics in 
the $s$ sector (the Birkhoff theorem). 
The energy vanishes on the constraint surface
(up to the surface term \cite{Regge:1974zd}). 

If $m\neq 0$, then  
varying ${\cal H}$ with respect to $N$ and $\beta$ gives 
\be                          \label{vars}
{\cal H}_0+m^2\frac{\partial \cal V}{\partial N}=0,~~~~~
{\cal H}_r+m^2\frac{\partial \cal V}{\partial \beta}=0.~~
\ee
For the generic potential \eqref{PF}, these relations can be resolved to 
express $N$ and $\beta$ in terms of $(q^i,p_k)$. No constraints 
then arise, so that all four phase space variables are independent 
and they describe 
two DOFs. One of them can be interpreted as the 
scalar polarization of the massive graviton, while the second one 
must be the ghost. 
Inserting $N(q^i,p_k)$ and $\beta(q^i,p_k)$ into ${\cal H}$, 
the result is not positive definite. In particular, the kinetic part of the 
energy associated with the momenta can be negative and
arbitrarily large,
which produces ghost instability 
\cite{Boulware:1973my}. 

In the dRGT theory, one has 
\be                     \label{V}
{\cal V}=\frac{NR^2P_0}{\Delta}
+\frac{R^2 P_1}{\Delta}\,\sqrt{(\Delta N+1)^2-\beta^2}
+R^2P_2\,, 
\ee
with 
$
P_n=b_{n}+2b_{n+1}\,{r}/{R}+b_{n+2}\,{r^2}/{R^2}.
$
The second relation in \eqref{vars} then determines the shift $\beta$,
\be                    \label{nu}     
\beta=(N\Delta+1)\frac{\Delta{\cal H}_r}{Y}\,,
\ee
where 
$
Y\equiv \sqrt{(\Delta {\cal H}_r)^2+({m^2}R^2P_1)^2}\,.
$
Inserting this into the first relation in 
\eqref{vars} does not, however, determine the lapse $N$,
but gives a constraint,  
\be                              \label{C}
{\cal C}\equiv{\cal H}_0+Y+{m^2}\frac{R^2P_0}{\Delta}=0.
\ee
Inserting 
\eqref{nu} into \eqref{Hs}
yields 
$
{\cal H}={\cal E}+N{\cal C},
$
with 
\bea                            \label{En}
{\cal E}&=& \frac{Y}{\Delta}+{m^2}R^2 P_2\,,
\eea
so that 
varying ${\cal H}$ with respect to 
$N$ reproduces the constraint equation 
${\cal C}=0$ once again. 
Since the constraint should be preserved in time, its Poisson bracket 
$\{,\}_{\rm PB}$ \cite{Hassan:2011ea,*Kluson:2012wf,*Comelli:2012vz}
with 
the Hamiltonian 
$
{H}=\int_0^\infty {\cal H}\,dr
$
should vanish. 
One can check 
that 
$
\{{\cal C}(r_1),{\cal C}(r_2)\}_{\rm PB}=0;
$ 
therefore 
$
{\cal S}\equiv \{{\cal C},{H}\}_{\rm PB}=0
$
is a new constraint since 
the term proportional to $N$ 
drops out of the bracket.
Explicitly, 
\bea                \label{S}
{\cal S}&=&
\frac{m^4R^2P_1^2}{2Y}\,
(\Delta\pii_\Delta+R\,\pii_R)
-Y\left(\frac{\Delta{\cal H}_r}{Y}\right)^\prime
\nonumber \\
&-&
\frac{\Delta^2\pii_\Delta}{2R}
\left\{
\frac{m^4}{2\Delta Y}\,
\partial_R (R^4P_1^2)+
m^2\partial_R (R^2P_2)
\right\} \nonumber \\
&-&
\frac{m^2{\cal H}_r}{Y}
\left\{
\Delta^2\left(R^2P_2\right)^\prime 
+R^2\partial_r(P_0-\Delta^2 P_2)
\right\}\,,
\eea
where $^\prime\equiv d/dr$ and
$\partial_R$ and $\partial_r$ are the partial 
derivatives with respect to $R$ and $r$. 
It is worth noting that  the 
two constraints have been known up to now only implicitly 
\cite{Hassan:2011ea,*Kluson:2012wf,*Comelli:2012vz}, whereas 
Eqs.\eqref{C} and \eqref{S} provide explicit expressions 
for any values of the parameters $b_k$. 
Requiring further that 
$\{{\cal S},{H}\}_{\rm PB}=0$ yields  
an equation for $N$ because 
$
\{{\cal C},{\cal S}\}_{\rm PB}\neq 0,
$
the 
term containing $N$ does not drop out. 

Since the constraints are second class, they remove one of the two DOFs
in the $s$ sector.
It is not immediately obvious which one is removed, 
but if the energy is positive, it follows that 
it is the ghost which is removed.

The energy is
$
E=\int_0^\infty {\cal E}\,
dr\,,
$
where the arguments of ${\cal E}$ 
should 
fulfill two constraint equations:
${\cal C}=0$ and 
${\cal S}=0$. 
These are nonlinear differential equations 
whose solutions $\Delta, R, \pii_\Delta$, and $\pii_R$ 
can be viewed as initial data for the Cauchy 
problem at the moment $t=0$. 
These equations are complicated,  
but they simplify in some cases. 

\section{Weak field limit}
In flat space, where
$N=\Delta=1$, $R=r$, and  $\beta=\pii_\Delta=\pii_R=0$, 
one has 
${\cal C}={\cal S}={\cal E}=0$.
If deviations
from flat space are small, then 
${\cal C}={\cal C}_{\rm FP}+\ldots$ and  
${\cal S}={\cal S}_{\rm FP}+\ldots$
(where the dots denote higher order terms), while the
leading terms  are, with $\delta\equiv \Delta-1$
and  $\rho\equiv R-r$,  
\bea                          \label{C_FP}
{\cal C}_{\rm FP}=
(2r(\delta+\rho^\prime))^\prime+
m^2(r^2\delta -2r\rho), \nonumber \\
{\cal S}_{\rm FP}=
\frac{m^2}{2}\,(r\pii_R-\pii_\Delta)-
(\pii_\Delta^{\prime}+\pii_R)^\prime.
\eea
If
$\nu\equiv N-1$, then
${\cal H}={\cal E}_{\rm FP}
+\nuu\,{\cal C}_{\rm FP}+\ldots$, where 
${\cal E}_{\rm FP}$ 
is  the leading  (up to a total derivative)
part of ${\cal E}+{\cal C}$,  
\bea                         \label{E_FP}
{\cal E}_{\rm FP}&=& 
\frac{\pii_\Delta^2}{4r^2}+\frac{\pii_\Delta \pii_R}{2r}
+\frac{(\pii_\Delta^\prime +\pii_R )^2}{m^2 r^2}\,\nonumber \\
&+&2\rho\,\delta^\prime-\rho^{\prime 2}-\delta^2
+m^2(
2r\delta\rho-\rho^2). 
\eea 
These are the FP  constraints and the energy density. 
The constraints are solved by 
\bea                          \label{FP_c}
\delta&=&-\rho^\prime+\frac{Q^\prime}{r^2},~~~~~~
\rho=\frac{Q}{r^2}+\frac{2Q^\prime}{m^2r^3}, \nonumber \\
p_R&=&-p_\Delta^\prime+\frac{F^\prime}{r},~~~~~~
p_\Delta=\frac{F}{r}-\frac{2F^\prime}{m^2r^2}\,
\eea 
for arbitrary $Q(r)$ and $F(r)$. 
Inserting this into \eqref{E_FP} gives
\be                      \label{T-FP}
{\cal E}_{\rm FP}=
\frac{3}{r^4}\left(
Q^{\prime 2}+m^2 Q^2+\frac{F^2}{4}
\right)+\left(\ldots \right)^\prime.
\ee
Since the fields should be weak, all of the above functions should be 
bounded, which imposes certain fall-off conditions on $Q$ and $F$ at the origin and 
at infinity. These conditions imply that the total 
derivative term in \eqref{T-FP}
vanishes upon integration; therefore the energy is positive.

\section {Momentum sector}
Let us assume that 
the three-metric is flat, so that $\Delta=1$ and $R=r$,   
while the momenta $\pii_\Delta$ and $\pii_R$ are not 
necessarily small and satisfy the two constraint equations
\eqref{C} and \eqref{S}.   
Introducing the dimensionless 
radial coordinate $x=mr$ with $^\prime\equiv d/dx$, 
expressing the two momenta in terms of two new function, $z$ and $f$, as 
\be 
\pii_\Delta=\frac{\sqrt{xz}}{m}\,,~~~~
\pii_R=-\frac{(xz+4x^4 f)}{2x\sqrt{xz}}\,,
\ee 
the constraints reduce to two equations, 
\bea                          \label{zf}
z^\prime &=&4\,x^2f+
2x\sqrt{xz}\,F\,, \\
f^\prime &=&\frac{
4\,(1-c_3)\,zf
-4x^3f-3z
}{4x\sqrt{xz}}\,F
-\frac2x\,F^2\,,\nonumber 
\eea
with $F=\pm\sqrt{f(f+2)}$, 
while the energy density 
$
{\cal E}=x^2f\,.
$
Since $F^2=f(f+2)\geq 0$, this implies that either one has 
$f\geq 0$ or $f\leq -2$, which 
determines two different solution branches whose energy is either
non-negative or strictly negative. 
There can be no interpolation 
between the branches, since this would require crossing the region of 
forbidden values of $f$.   

A simple solution from the first branch is $f=0$ and $z=z_0$, whose energy is 
zero. It reduces to the flat space configuration for $z_0=0$. 
If the solutions of Eq.\eqref{zf} are to describe initial values 
for perturbations around flat space,
then 
they should correspond to smooth deformations of the latter,
and this selects the $f\geq 0$ branch. Therefore, 
the energy for perturbations around flat space 
is positive.  

 A simple solution from the second branch is 
$f=-2$ and $z=\frac{8}{3}\,(x_{\rm max}^3-x^3)$, 
where $x_{\rm max}$ is an integration constant. 
Since $z$ should be positive, the solution exists only
for $x\leq x_{\rm max}$, with 
the total energy 
$E=\int{\cal E}\,dr=-\frac{2}{3m}\,x_{\rm max}^3$.
One can construct more general negative energy solutions of Eq.\eqref{zf}
numerically. They exist 
only within finite intervals of $x$ at whose ends one has 
either $f\to -\infty$ or $z\to 0$.
Such solutions cannot describe regular initial data and they  
belong to the disjoint from flat space branch.

Summarizing, the energy of smooth excitations over the flat space is positive.
It can also be negative,
but only in a sector disconnected from flat space;
therefore this cannot lead to the ghost instability of the latter.

\section{Metric sector}
Let us now set the momenta to zero, 
$\pii_\Delta=\pii_R=0$, without fixing $\Delta$ and $R$.   
This solves the second constraint, 
$
{\cal S}=0. 
$
Denoting $h=R/r$ and  $g=\Delta h$, using again   
$x=mr$ with $^\prime\equiv d/dx$, and setting 
$c_3=c_4=0$ for simplicity, the first constraint reduces to 
\bea                           \label{eqC0}
&&h^{\prime\prime}
+\frac2x\,h^\prime-
\frac{h^{\prime 2}}{2h}+\frac{(xh)^\prime g^\prime}{xg}
-\frac{h(1-g^2)}{2x^2g^2} \nonumber \\
&-&\frac{h(2-3h)}{2g} 
-\frac{h(1-6h+6h^2)}{2g^2}
=0,
\eea
and the energy density 
\bea                  \label{eqE}
{\cal E}&=&\frac{x^2h^2(3h-g-2)}{g}\,.
\eea
The constraint is solved  by setting $g=qh/(xh)^\prime$ and 
\bea                           \label{eqC}
Q&=&xh(1-q^2)+x^3 h(h-1)(2h-1), \nonumber \\
Q^\prime&=&x^2 h(q-1)(3h-2), 
\eea
with any $Q(x)$. Having chosen
$Q$, these algebraic equations can be resolved   
with respect to $h$ and $q$.
Even though the second constraint is trivially satisfied, 
its stability condition, 
 $\{{\cal S},H\}_{\rm PB}=0$, is nontrivial,
$
{\cal A}N-{\cal B}=0\,,
$ 
 with 
\bea
{\cal A}&=&q(\alpha_1+\alpha_2)+2(q-1)^2(27h^2-18h+4)(xh)^\prime\nonumber   \\
&-&6x^2h(3h-1)(3h-2)(4h-3)(xh)^\prime, \\
{\cal B}&=&(\alpha_2-\alpha_1)(xh)^\prime+8h^2q(q-1)^2
+6h^2x^2q(3h-2)^2\,,\nonumber 
\eea
where 
$\alpha_1=3x^2h(3h-2)(13h^2-12h+2)$ and  
$\alpha_2=2h(q-1)^2(9h-2)$.
Thus the lapse function is $N={\cal B}/{\cal A}$,
while the shift function obtained from Eq.\eqref{nu} is $\beta=0$.  

The three-metric will be regular 
and asymptotically flat 
if $h$ and $q$ are smooth and fulfill the  boundary conditions 
\be                  \label{bc}
h_0\leftarrow h \to 1,~~
1\leftarrow q \to 1~~~~~\mbox{for}~~
0\leftarrow x \to \infty,
\ee
with $h_0>0$. The simplest solutions of the constraint 
are obtained by setting in \eqref{eqC} $Q=0$,
which implies that $q=1$, but yields two different solutions for $h$: 
\be                              \label{h}
h(x)=\left\{1,\frac12\right\}~~~~\Rightarrow
~~~~~{\cal E}(x)=\left\{0,-\frac38\,x^2\right\}.
\ee  
These fulfill also 
the Hamilton equations, $\dot{q}_k=\{q_k,H\}_{\rm PB}=0$ and  
$\dot{p}_k=\{p_k,H\}_{\rm PB}=0$, and give rise 
to two different branches of more general solutions of the constraint.

\subsection{Normal branch}
\begin{figure}[h]
\hbox to \linewidth{ \hss
	
	\resizebox{7cm}{5.5cm}
	{\includegraphics{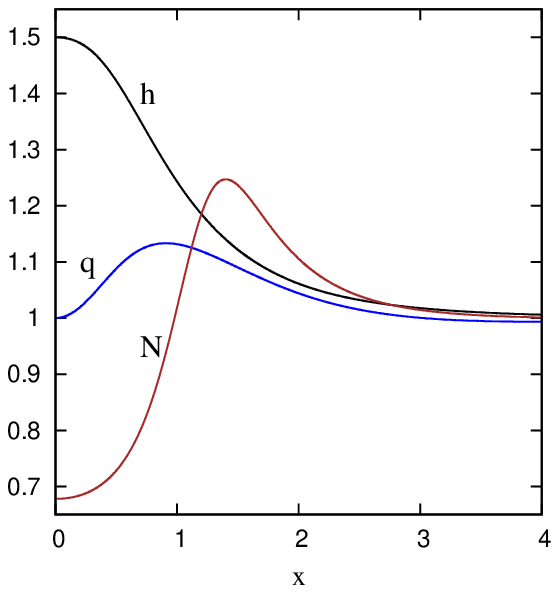}}
\hspace{1mm}
	\resizebox{7cm}{5.5cm}{\includegraphics{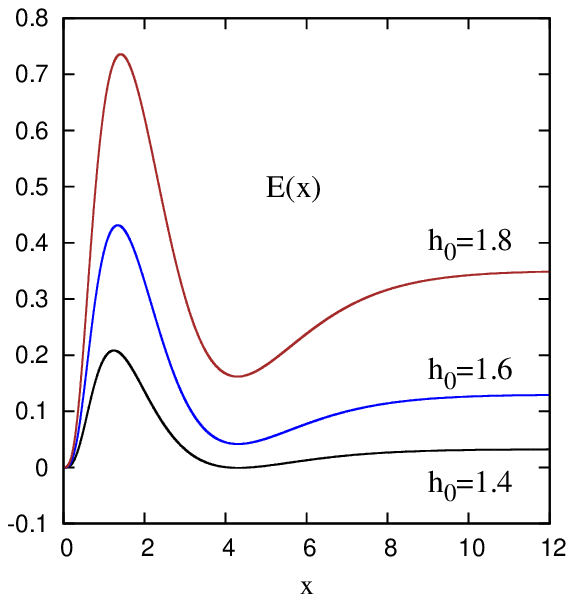}}
	
\hspace{1mm}
\hss}
\caption{{\protect\small 
Profiles of $h,q,N$,  and $E(x)$  for 
the positive energy solutions with $Q=kr^5/(1+x^2 e^x)$. 
 }}%
\label{Fig1}
\end{figure}

For the $h=1$ solution in \eqref{h}, one has  
$N=1$, so that the four-metric is flat, $ds_g^2=\,ds_f^2$, 
and the energy is zero.   
For deformations of this solution, one has  $h=h_0+{\cal O}(x^2)$ and 
$q=1+{\cal O}(x^2)$ for small $x$, in which case 
Eqs.\eqref{eqC0} and \eqref{eqC} require that 
$Q=kr^5+{\cal O}(x^7)$ 
with $k=0.1\times h_0(2h_0-1)(h_0-1)(3h_0-2)$. 
This suggests that one can choose 
$Q=kr^5/(1+Ax^2 e^x)$, and resolving Eq.\eqref{eqC} with respect to 
$h$ and $q$ then gives the global solutions shown in Fig.\ref{Fig1}. 

These solutions are smooth and globally regular; they describe 
initial metric deformations of flat space. Interestingly, 
the energy (expressed in $1/m$ units) contained in the sphere or radius $x$, 
$
E(x)=\int_0^x {\cal E}\, dx, 
$
can be negative for small $x$ (if $h_0<1$), but the total energy 
$E(\infty)$ is always positive and grows 
when $|h_0-1|$ increases. 
As a result, the energy is positive for smooth, asymptotically flat 
fields, so that the 
positivity of their energy in the weak field limit 
holds in the fully nonlinear theory as well.

\subsection{Tachyon branch}
For the $h=\frac12$ solution in \eqref{h}, 
one has $N=\frac12$, so that the  
metrics are proportional, $ds_g^2=\frac14\,ds_f^2$.
Even though they are both flat, 
this solution is quite different from flat space
since one now has  $E(x)=-x^3/8$, which corresponds to the constant
negative energy density. The total energy is negative and infinite. 
Considering small fluctuations around this background, 
the corresponding Fierz-Pauli mass is $m_{\rm FP}^2=-\frac12\, m^2$ 
[as is seen by linearizing the constraints and comparing with \eqref{C_FP}],
hence gravitons become tachyons, which can be viewed as an indication  
of the presence of the ghost. 

One can also construct more general solutions of this type 
by setting in \eqref{bc} $h_0\approx \frac12$, in which case 
$h(x)\to\frac12$ as $x\to\infty$. 	 
The energy is always 
negative and infinite.  
However, none of these solutions  
fulfill the boundary condition \eqref{bc}. Since they 
are not asymptotically flat, they cannot affect 
the stability of flat space.

\subsection{Tachyon bubbles}
\begin{figure}[h]
\hbox to \linewidth{ \hss

	\resizebox{7cm}{5.5cm}
	{\includegraphics{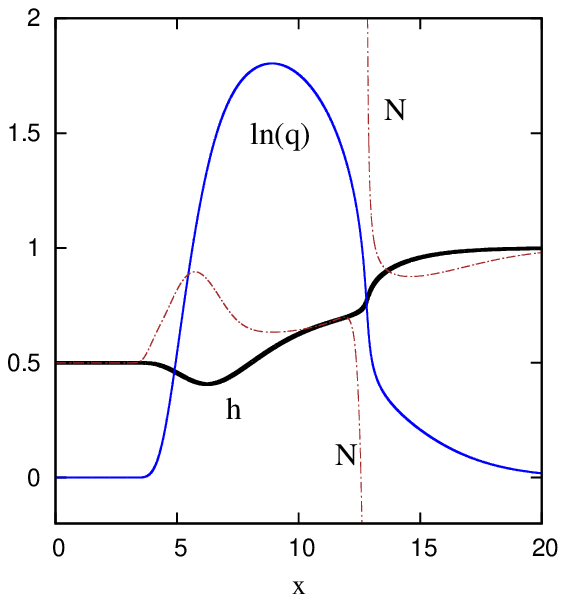}}
\hspace{1mm}
	\resizebox{7cm}{5.5cm}{\includegraphics{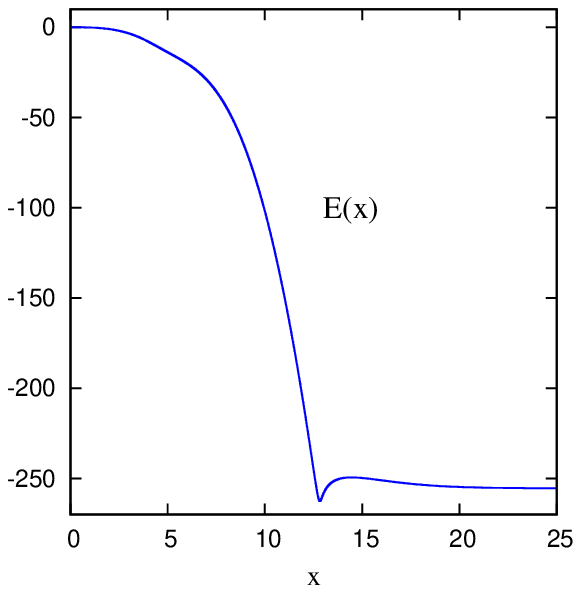}}
	
\hspace{1mm}

\hss}

\caption{{\protect\small Profiles of $h$ (thick line), 
$\ln(q)$, $N$, and $E(x)$ for the tachyon bubble solution
with  $Q=-\Theta(x-3)(x-3)^8 e^{-x}$.
 }}%
\label{Fig2}
\end{figure}

There are also asymptotically flat solutions whose energy is finite and negative. 
They can be obtained by 
choosing in \eqref{eqC}  
$Q=A\,\Theta(x-x_0)(x-x_0)^p e^{-x}$, where $\Theta(x)$ is the step function. 
This enforces for $h$ a kink-type behavior, so that 
$h=\frac12$ for $x<x_0$, but $h$ increases for $x>x_0$  and approaches 
unity as $x\to\infty$ (see Fig.\ref{Fig2}). Solutions thus approach 
the flat space at infinity, but they contain a bubble of the tachyon phase 
in a finite region.  If $x_0$ is large, then the energy 
$E\propto -x_0^3$ (see Fig.\ref{Fig2}).
The existence of such solutions is embarrassing, 
since it suggests  that the flat space
could be unstable with respect to decay into bubbles. 

However, 
 a closer inspection reveals that the lapse function $N$ for the bubbles 
is singular. 
Indeed, one has $N={\cal B}/{\cal A}$, but both ${\cal A}$ and ${\cal B}$
have opposite signs for $h=\frac12$ and  $h=1$; hence they vanish 
at least once as $h$ interpolates between $\frac12$ and $1$. 
One can show that they cannot vanish simultaneously; therefore 
$N$ must have at least one zero and a pole, as shown in Fig.\ref{Fig2}. 
Since $N$ enters the Hamilton equations 
$\dot{p}_k=\{p_k,H\}_{\rm PB}$, the time derivative of the momenta 
diverges where $N$ has pole(s). Therefore, the bubble solutions do not 
describe regular initial data, so that they cannot provoke instability 
of flat space.  
The tachyon bubbles can be obtained also for other values of the 
theory parameters $c_3$ and $c_4$, but their lapse 
function $N$ is always found to be singular.

\section{Stability of the theory}
To recapitulate, the above results indicate that 
the energy in the dRGT theory is positive  for 
globally regular and asymptotically flat fields. 
The energy can also be negative and even unbounded from below,
but in all studied cases the corresponding solutions are found to be either 
not asymptotically flat or not global or singular. 
They cannot describe initial data for a decay of the flat space. 
Therefore, one is bound to conclude that there is evidence for the 
stability of flat space,  
despite the existence of the negative energies.

One can provide the following interpretation.  
 Globally regular and asymptotically flat fields constitute the 
``physical sector'' of the theory where the energy is positive and the 
ghost is absent/bound. One may hope that a positive energy theorem
can be proven in this case. 
As for the negative energy states, 
they belong to disjoint sectors and cannot communicate with the physical sectors
since they are singular. 
Therefore, even though the negative energies can be viewed as an unphysical 
feature, they are harmless because they decouple.  

One may wonder if these classical arguments could be extended to show that the
 physical sector is protected  against quantum 
corrections. Let us estimate  the 
height of the potential barrier between the different  sectors. This can be 
done by computing the energy for  interpolating sequences of fields. For example, 
fields which fulfill the constraint and  
satisfy the boundary conditions  \eqref{bc} will interpolate 
between the normal and tachyon branches when the 
parameter $h_0$ in \eqref{bc} decreases from $1$
to $1/2$. It turns out that when $h_0$ starts decreasing, the energy rapidly grows,
since the function $g$ in the denominator in \eqref{eqE} develops a minimum
(the lapse function $N$ then typically shows several  poles). As $h_0$ continues 
to decrease, the energy passes through a simple pole and then 
approaches a finite negative value 
when $h_0$ tends to $1/2$. Therefore, the potential barrier between the two 
branches is infinitely high. 

However, it is still possible that the barrier height could be made finite via minimizing 
the energy with respect to the function $Q(x)$ in \eqref{eqC}. 
Let us suppose that the minimal barrier height indeed has a finite value, $E_{b}>0$. 
This value cannot be arbitrarily small, since
when one starts deviating from flat space the energy 
grows, because the Fierz-Pauli energy is positive. Therefore, 
the energy  can show a maximum and start decreasing only when the nonlinear
effects become essential, but by this moment it should already assume 
a finite value. The dimensionful energy is obtained by dividing by the graviton 
mass, $E_{\rm b}/m$, and since $m$ is extremely small, the energy will be extremely large, 
of the order of the total 
energy contained in our Universe. As a result, even if the potential barrier between
different sectors was finite, it would be 
cosmologically large, implying that the physical sector should actually  be stable
both classically and quantum mechanically.   

It is also worth noting that within the bigravity generalization of the dRGT 
theory   where both metrics are dynamical \cite{Hassan:2011zd},
the tachyon vacuum in \eqref{h} is no longer a solution, as it does not fulfill the equations for the second metric. Therefore, since there are less negative energy solutions, 
it seems that the positivity of the energy should be easier to demonstrate
when both metrics are dynamical. Similarly, including a matter source should also have a stabilizing 
effect since the energy of the physical matter is expected to be positive.

The decoupling of the negative energies suggests that the other seemingly unphysical features 
of the dRGT theory, such as the superluminality 
\cite{Deser:2012qx,*Deser:2013rxa,*Deser:2013eua,*Deser:2013qza,*Deser:2013gpa}, 
may also decouple. In fact, it has not been shown that the superluminality should inevitably  
develop starting from any smooth initial data. On the contrary, one could expect
different unphysical features to come up together, so that  
superluminal fields can be expected to have negative energies. 
But then they should decouple. Although not a proof, 
this indicates  that the 
superluminality 
could perhaps relate only to the unphysical sectors, in which case it would be harmless. 

\section{Acknowledgments}
It is a pleasure to acknowledge discussions with Claudia de Rham, 
Andrew Tolley, and 
Cedric Deffayet and interesting 
remarks of Eugeny Babichev.   
This work was partly supported by the Russian Government Program of Competitive Growth 
of the Kazan Federal University.




%

\end{document}